# A citation history of measurements of Newton's constant of Gravity

**Katelyn Horstman**

Department of Physics and Astronomy, University of California Los Angeles – Los Angeles, CA 90024, USA

E-mail: katelynhorstman@gmail.com

**Virginia Trimble**

Department of Physics and Astronomy, University of California Irvine – Irvine, CA 92697, USA

E-mail: vtrimble@astro.umd.edu

November 9, 2018

## Abstract

We created and analyzed a citation history of papers covering measurements of Newton's constant of gravity from 1686 to 2016. Interest concerning the true value of the gravitational constant was most intense in the late 90s to early 2000s and is gaining traction again in the present. Another network consisting of the same papers was created using citations from databases to display the prominence of papers on Newton's constant in the wider scientific community. In general, papers that were important in one network remained important in the other while papers that had little importance in one network remained unimportant in the other.

The US contributes the most to literature on the topic both in where journals were published and where the work was done; however, many other countries, such as China, Russia, France, Germany, Switzerland, and the UK also provide many papers on Newton's G. Work done within certain countries tends to be considered more important and cited more often within that country. Recent efforts promoting international collaboration may have an impact on this trend.

## Introduction

In 1789, Henry Cavendish was to use the first torsion balance to measure the strength of gravitational interaction. After his initial success, many others explored methods of measuring the gravitational constant, either by modifying his experiment or seeking better ones. However, even with experimentation and concern for finding the true value of the constant over the past few centuries, Newton's constant remains elusive. It is by far the least well known of the fundamental constants of physics. Measurements from a handful of differing experiments – even more recent measurements – have discrepancies of up to .05%, leading to large uncertainty in the true value of the constant. Researchers agree on at most the first three significant figures, or that $G = 6.67 \times 10^{-11} \text{N} \times \text{m}^2/\text{kg}^2$.

The uncertainty associated with Newton's gravitational constant is concerning because gravity is the most easily experienced and most recognizable fundamental force of nature. A very precise value could be a hint of some physics beyond general relativity. It is notoriously tricky to measure for of three reasons. First, gravity is an extremely weak force compared to other measured forces such as electromagnetic or nuclear forces. For a proton and an electron, the ratio of the electromagnetic to the gravitational force is given by $\frac{F_{EM}}{F_G} = \frac{e^2}{4\pi\varepsilon_0 G m_p m_e} = 10^{40}$, where $e$ is charge on the electron, G is Newton's constant, $m_p$ is the mass of the proton, and $m_e$ is the mass of the electron. Second, G is not tightly coupled in with the constants of electromagnetism and quantum mechanics in the way that the mass and charge of an electron are. And third, all measurements of Newton's gravitational constant have had to be done very close to a large, interfering mass – Earth. In 2016, the National Science Foundation began a quest for a more precise value of "Big G." They convened a workshop at NIST (National Institute for Standards and Technology) to propose and discuss novel measurement techniques to resolve discrepancies and lead to a more precise value for G. One of us, Virginia Trimble, was a participant.

In this paper, we explore the relationship between papers searching for or discussing the importance of Newton's gravitational constant. Dating from Newton in 1686 to efforts in 2016, the citation map created shows some of the most important relationships between literature published on Newton's constant within the past few centuries and the resources those papers used for inspiration. The citation map helps identify which papers are the most influential within the community. We aim to discover whether members within this community are receiving the credit they deserve or if other members are overrepresented. We also intend to see if countries in which the work was done or journals in which the papers were published in affect the amount of influence a paper has.

Katelyn Horstman thought of looking for correlations with gender but discovered that the use of initials often made this impossible to determine and we suspect that there are very few women authors in this network.

Stephan Schlamminger, a physicist at the National Institute of Standards and Technology (NIST), kindly gave Virginia Trimble a flash drive consisting of the papers he deemed important concerning Newton's gravitational constant. She concluded that some sort of citation analysis, grouping publications and using citations from outside the group, would be useful. We created the citation analysis using around 220 of the roughly 300 original papers from the flash drive. The 80 or so papers not used did not help build the network as they either did not cite papers within the list or consisted of papers with no citations at all. We created another network consisting of the same papers, but each paper's graphical representation was based the number of times it was cited within the scientific community.

## Methods

We used an excel spreadsheet to document and cite the papers within the network. Each paper in the network either cited another paper within the network or was cited by another paper within the network. We formatted the sources and their citations to establish relationships between papers then uploaded the created relationships to a citation analysis program called Gephi. Gephi created a citation network graphically showing the relationship among papers.

Each paper was assigned a number, listed in the Appendix, and then ordered chronologically on the diagram. Lines connecting different numbers show which papers cited other papers and the direction is indicated with small arrows pointing toward the paper that was cited. The larger the number appears in the diagram, the more times that paper was cited. In addition, most frequently cited papers are purple, moderately cited papers are yellow, and least frequently cited papers are white. The network shown displays the relationship between various methods of calculating Newton's gravitational constant and other papers documenting its importance or history.

Data

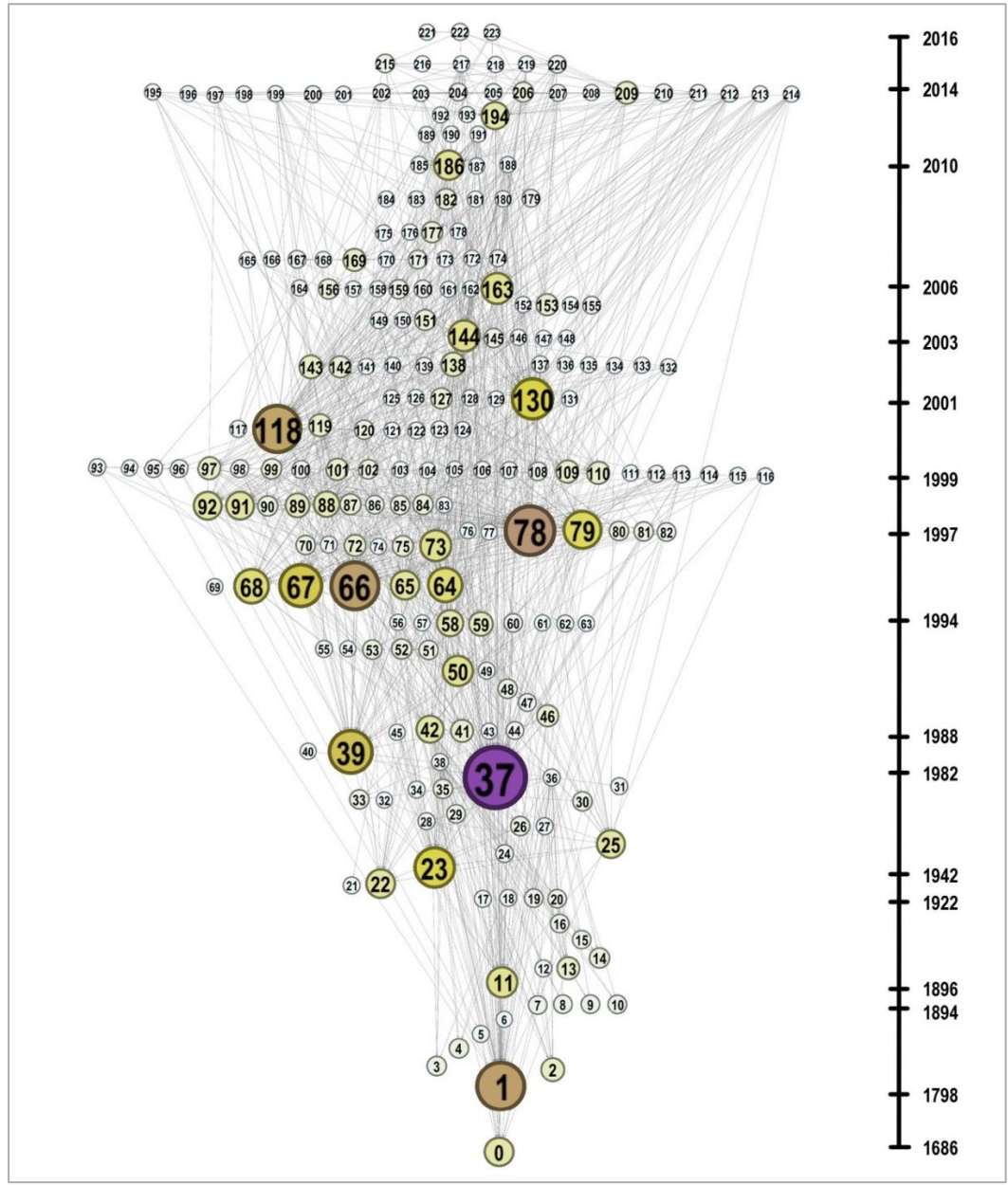

**Figure 1.** A citation network showing the relationship between papers influential in studying Newton's gravitational constant. Each line shows a connection between two papers and is numbered in Figure 12.

Figure 1 shows how many times each paper in the network was cited by another paper in the network. As predicted, recent papers cite older papers more frequently. In the late 90s to early

2000s showed a surge in notable papers concerning Newton's G. The next wave of papers occurs in 2014. Physicists from around the world met at the Royal Society in London and published many findings – perhaps helping renew the National Science Foundation's interest in finding a better value for Newton's gravitational constant.

Looking at the shape of the network itself, we found that the number of papers starts small, grows large in the middle, then gets small again when it approaches more recent times. The smaller number of papers published in 2016 could indicate not knowing which papers will be important in future instead of a lack of interest in the topic.

After reviewing the connections made in the network, we decided we wanted to investigate the importance of each paper in the greater scientific community. Using the Astrophysics Data System and the Web of Science databases, we used the number of times each paper was cited to resize and recolor the circles based on their importance in all fields. Figure 2 shows the results.

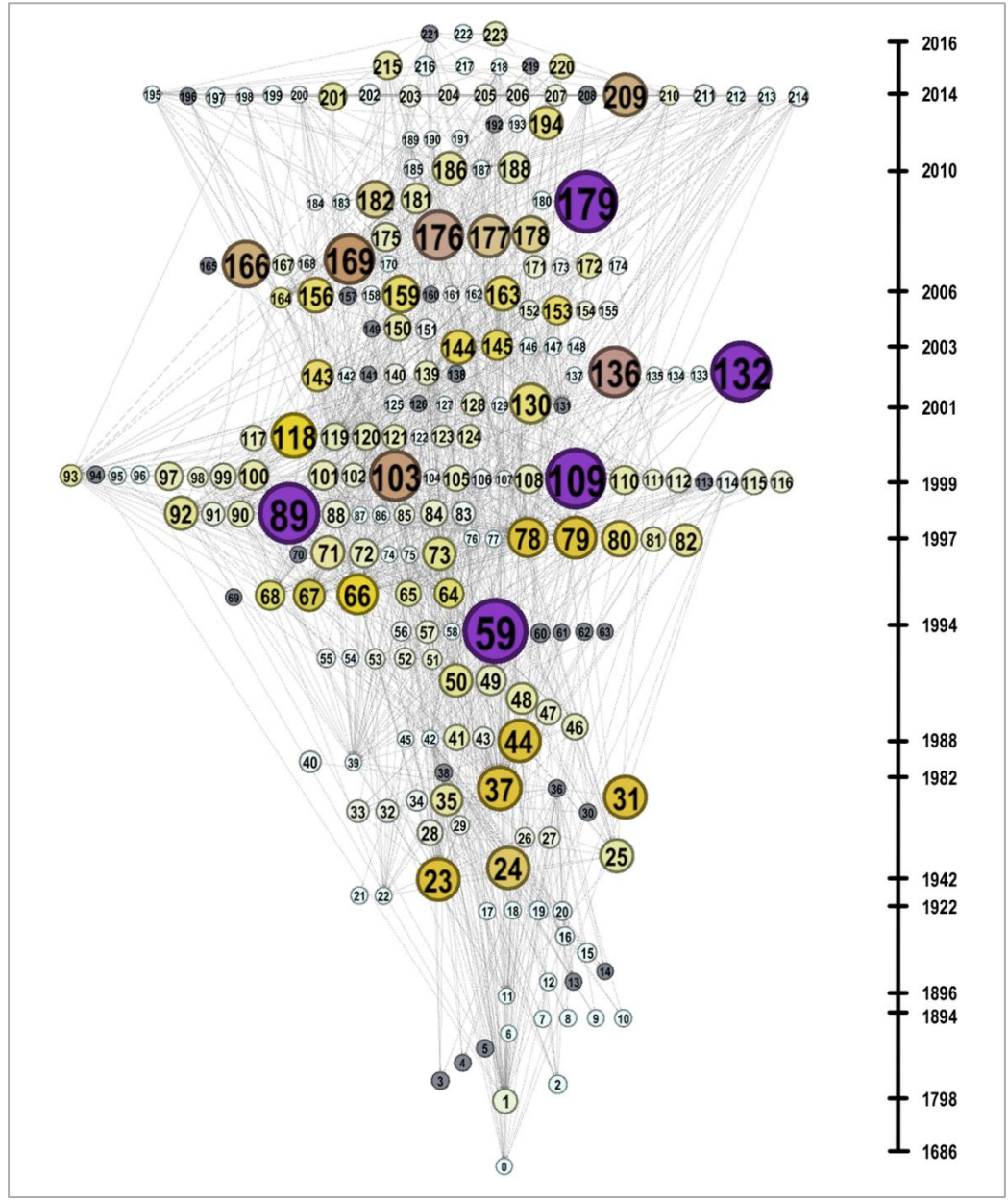

**Figure 2.** A citation network showing the relationship between papers influential in studying Newton's gravitational constant. Each paper's size and color are based on how often the paper was cited in the scientific community. The smaller, gray circles represent papers not found in a database. The original connections between papers from Figure 1 remain.

In general, papers that were important in Figure 1 remained important in Figure 2 and the papers that could not be found have little importance in both figures. Papers from almost all time periods remain important to scientists based on their citation history. A larger number of papers in Figure 2 have a substantial number of citations. Some of the larger circles in Figure 2 have more than 400 citations while the larger circles in Figure 1 have fewer than 100 citations due to the size of the network.

## Analysis

After creating the citation networks in Figures 1 and 2, we decided to look more deeply at the countries associated with the journals the papers were published in.

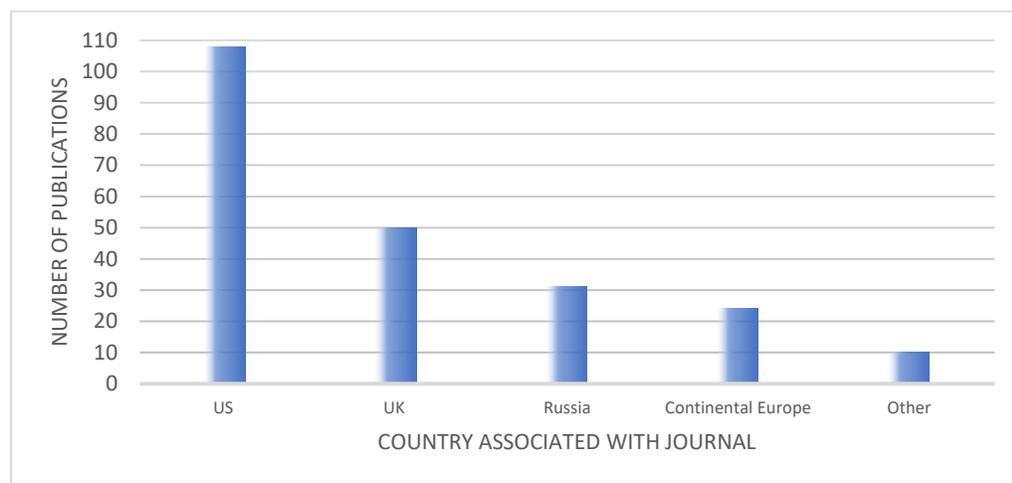

**Figure 3**. A bar graph representation of the total number of papers in the network published in different journals in the network. Each journal is associated with the US, UK, Russia, Continental Europe, or other countries not listed before.

Figure 3 shows the US journals dominate where scientists publish their results on Newton's gravitational constant. The UK, Russia, and countries in Continental Europe all have a moderate number of papers published in their respective journals while countries such as China have few papers published on Newton's G. However, the country associated with the journal is not always the same as the country where the work was done.

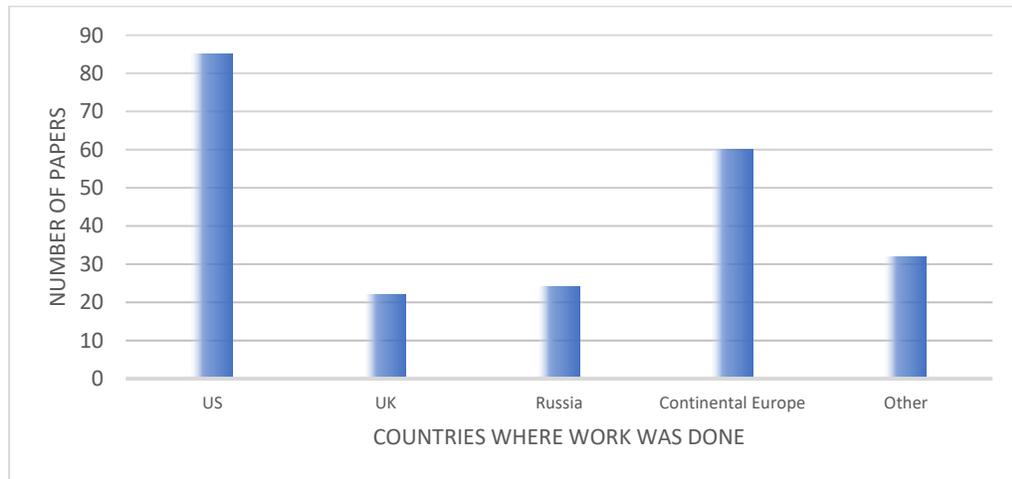

**Figure 4.** A bar graph showing countries where work was done versus the number of papers associated with that country in the network.

Figure 4 shows the slight discrepancy between countries where the work was done versus what country the journal is associated with after the paper was published. Although most of the work was done in the US, there are still a significant number of papers published in US journals even though the work was done in different countries. For quite a few papers, the work was done in either Continental Europe or other countries not listed in Figure 4, but the papers were published in US journals. Many papers where experiments were carried out in China published their result in either US or UK journals. We speculate this could be due to the regular interest the US and UK have had in Newton's constant starting in the early 1900s and continuing until the present.

World War II was another barrier when measuring Newton's gravitational constant. Figure 5 shows the country the work was done in before 1938, or the beginning of World War II, while Figure 6 shows the country the work was done in after 1938.

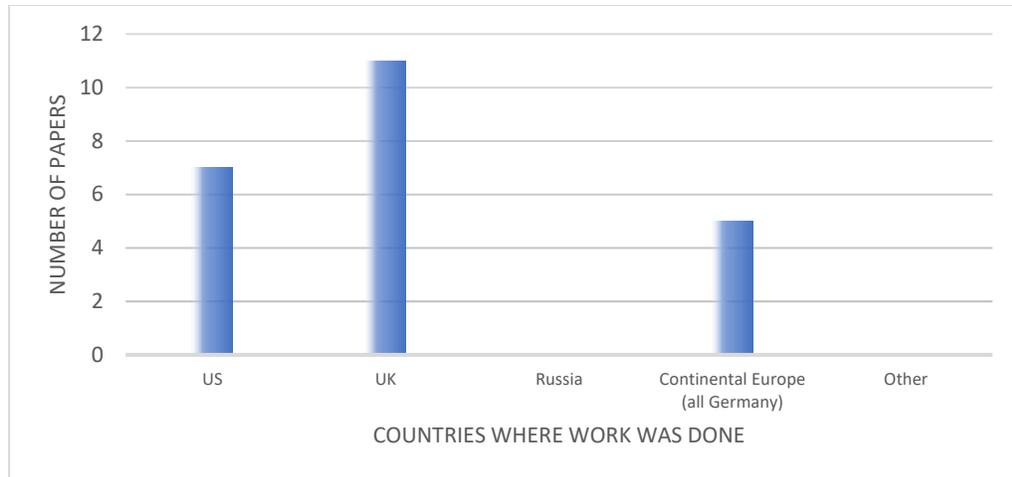

**Figure 5.** The number of papers published before 1938 versus where work was done.

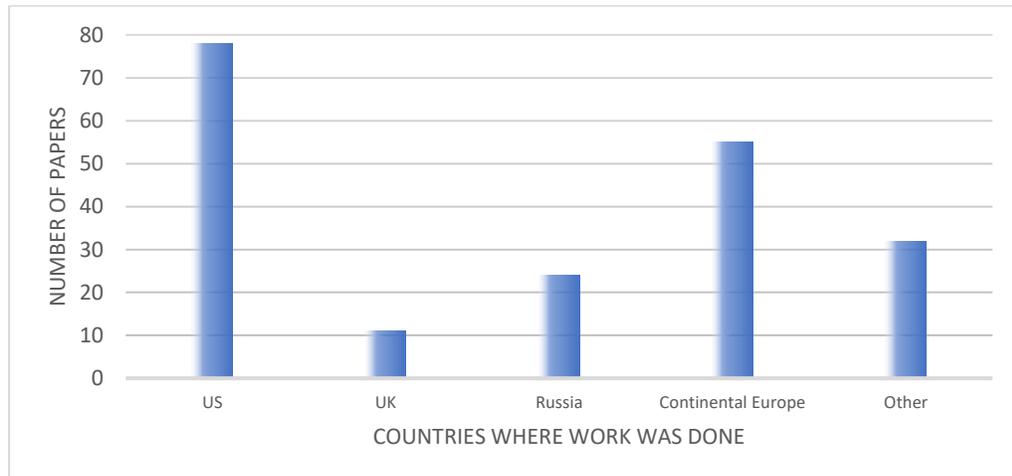

**Figure 6.** The number of papers published after 1938 versus where work was done.

Figures 5 and 6 show the overall increase in interest in Newton's constant after 1938. Before 1938, the UK, not the US, published the largest number of significant papers regarding G. However, after 1938, the US dominates where the work was done. Looking more closely at Continental Europe, all the papers published before 1938 described that the work was done in Germany. After 1938, Continental Europe showed a renewed interest in Newton's constant, but the work was mostly done in countries other than Germany, as illustrated by Figure 9. Russia and other countries not listed in Figures 5 and 6 did publish important works in physics before 1938 but did not seem to be concerned with the gravitational constant until after that time.

After looking at the countries where the work was done, we made smaller citation analysis clusters based on specific countries to look more in depth at the influences countries have on their own papers. We looked at countries that had a small cluster of papers to see if could discover anything intriguing about how countries influence their own work compared to how they influence global work.

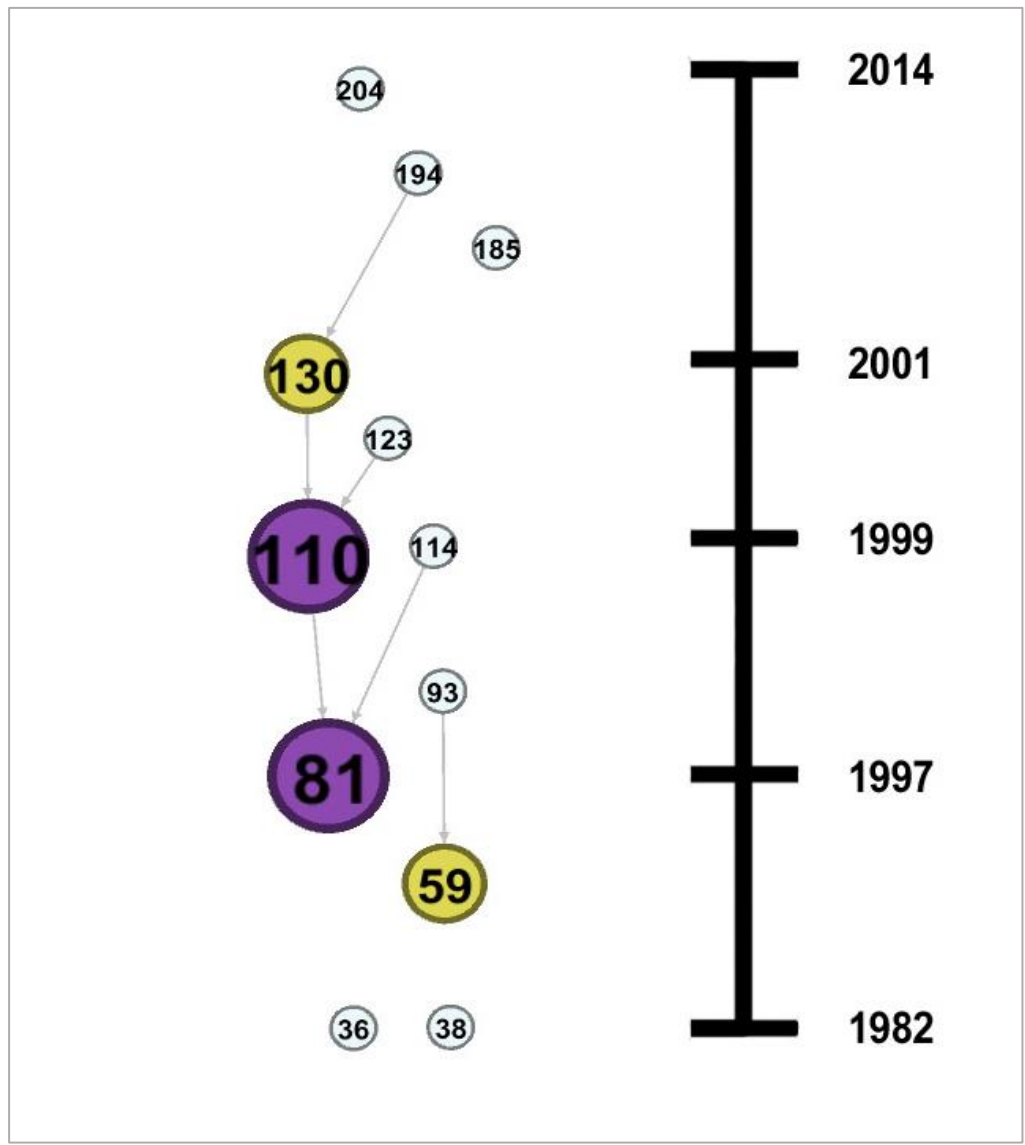

**Figure 7.** Small cluster of papers with work done in Russia.

Figure 7 shows the small cluster of papers where the work was done in Russia. Many papers in this network do not cite any other papers from Russia nor get cited by papers from Russia. The most important work done by Russia was in the late 90s to early 2000s which matches the overall importance of Russian papers in Figure 1 and 2.

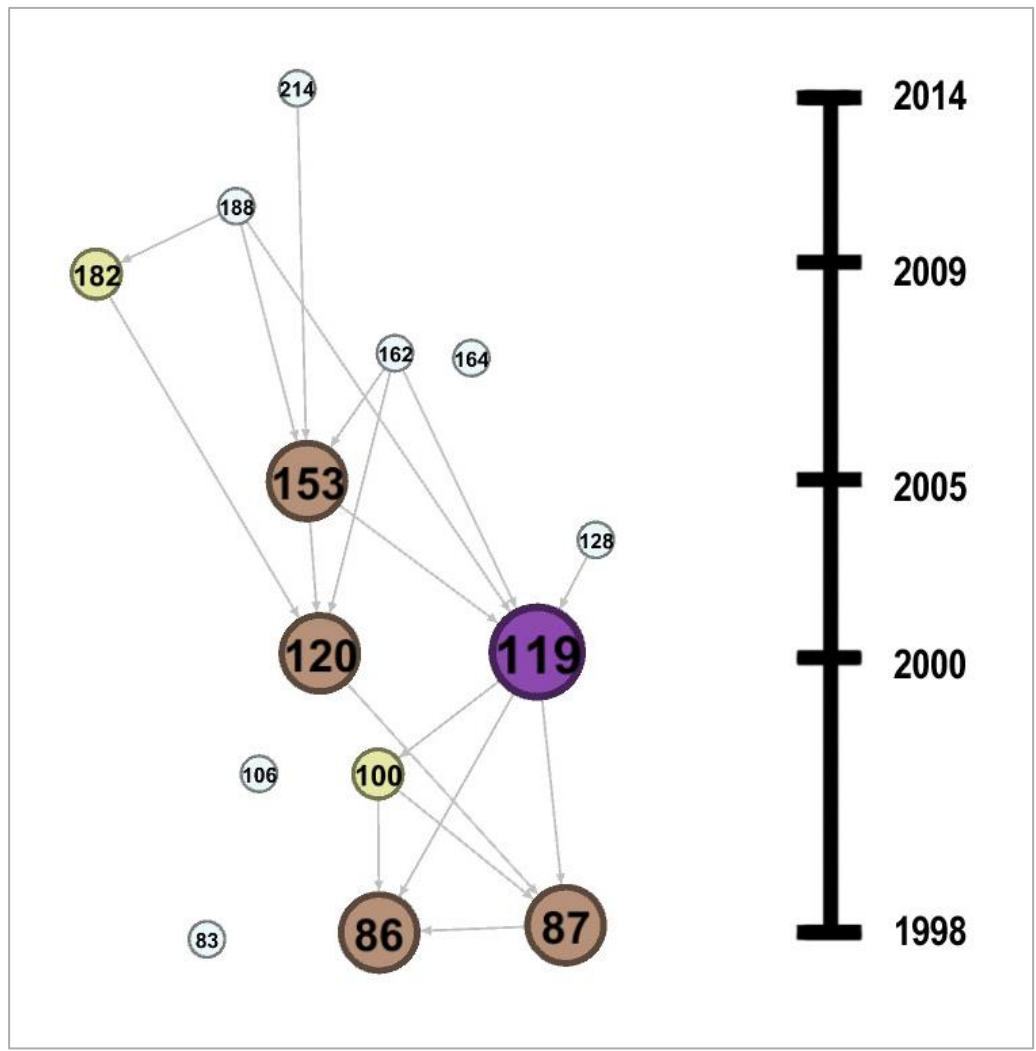

**Figure 8.** Small cluster of papers with work done in China.

Figure 8 shows a small cluster of papers where the work was done in China. A larger majority of papers from China cite work from within the country given the smaller network only consists of fifteen papers. Many Chinese authored papers also cite US and UK papers. The network shows the influence Chinese papers have on each other and the possible influence Chinese papers may have on determining how relevant papers from outside China by citing them.

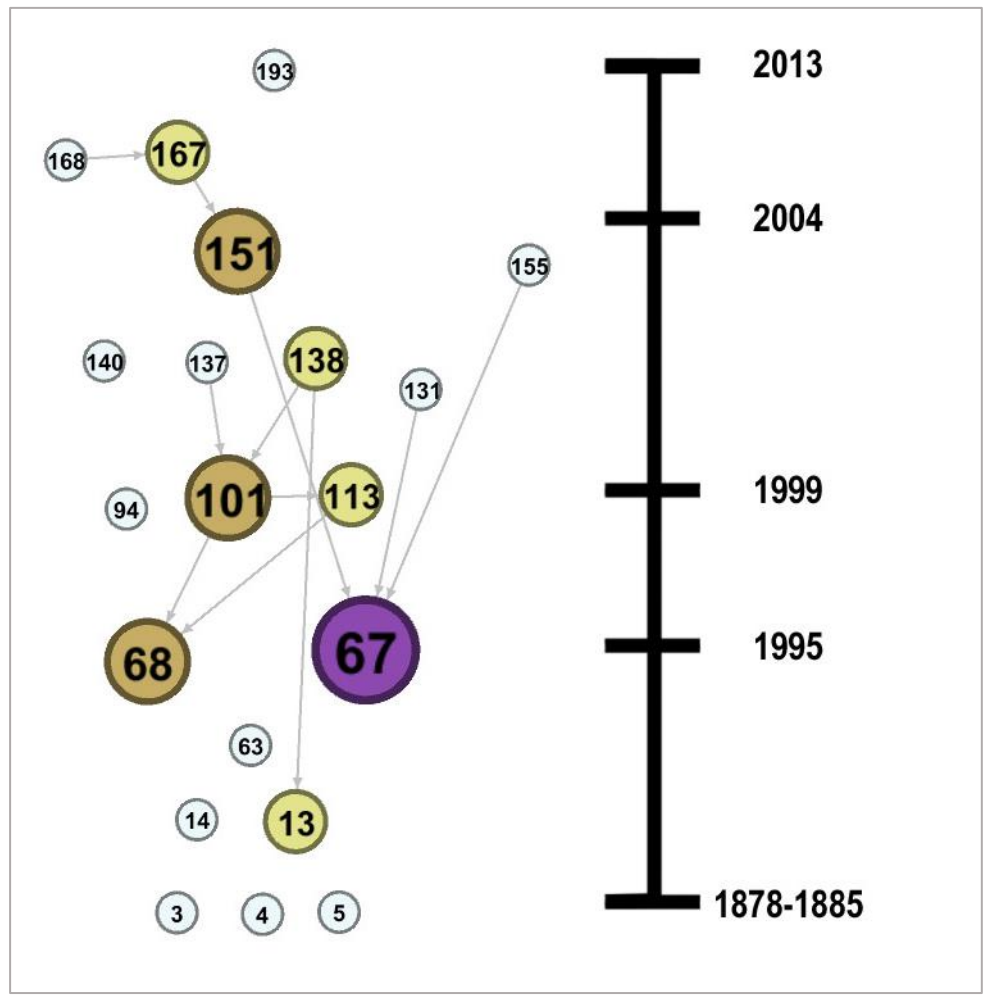

**Figure 9.** A small cluster of papers with work done in Germany.

Figure 9 shows a small cluster of papers where the work was done in Germany. A moderate number of papers cite each other within the network given its size. However, many of the older papers do not seem to be a foundation for work within the contemporary study of Newton's gravitational constant. As mentioned, a lot of work was done in Germany before 1938, but work on the constant did not really gain significance and pick up again until the mid-90s.

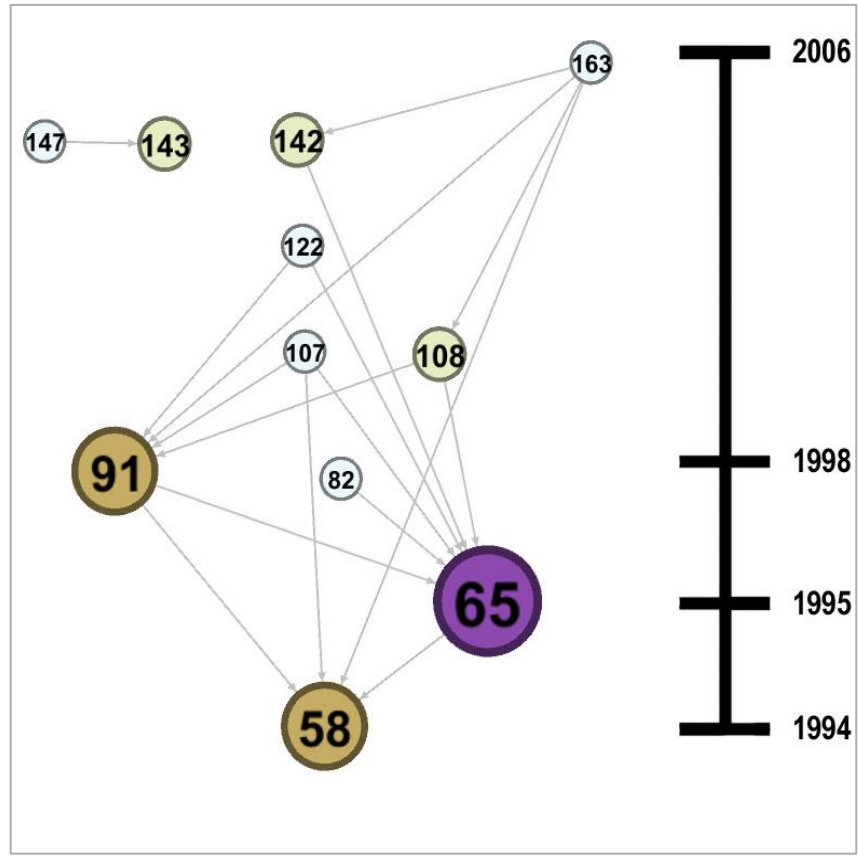

**Figure 10.** A small cluster of papers with work done in Switzerland.

Figure 10 shows a small cluster of papers where the work was done in Switzerland. The Swiss network seems to be the most tightly interlinked of the five networks that are broken down into smaller clusters. Every paper in the Swiss network cites or is cited by another paper within the network. Even though papers 58, 65, and 91 are crucial in the Swiss network, in Figures 1 and 2, they remain only moderately critical.

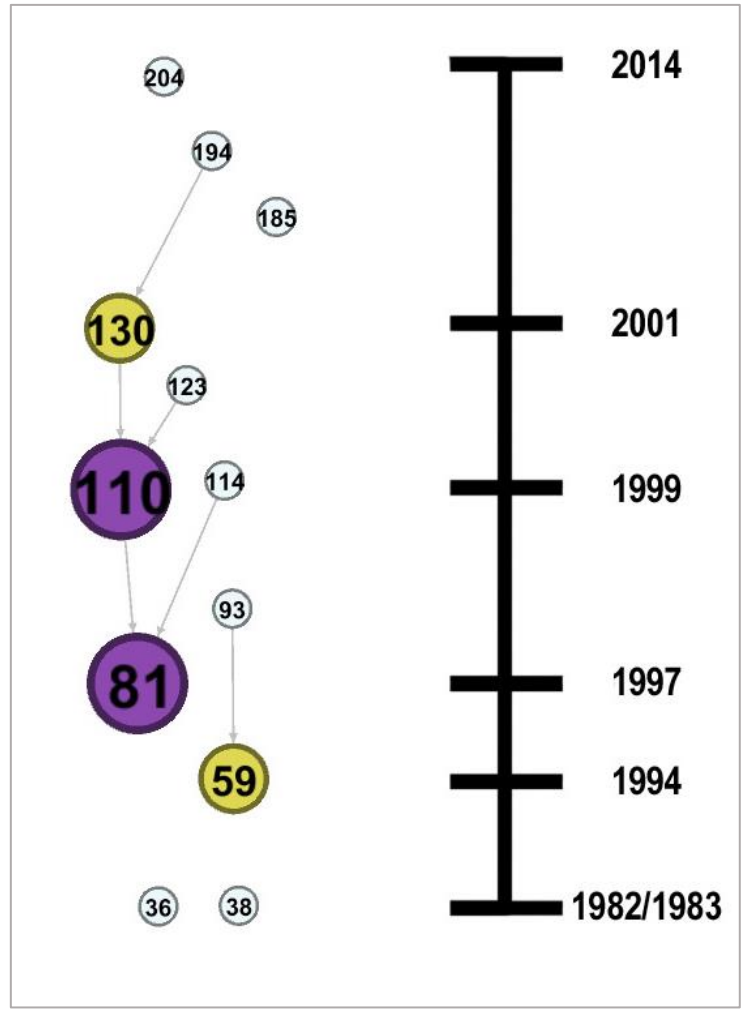

**Figure 11.** A small cluster of papers with work done in France.

Figure 11 shows a small cluster of papers where the work was done in France. A smaller than expected number of papers within the network cite other papers from France given the size of the network.

Conclusion

This paper discusses the implications of the overall trends of a citation analysis done on papers exploring measurements of Newton's gravitational constant. Work slowed down on finding the true value of Newton's G after the early 2000s, but seems to pick back up again beginning in 2014. Looking more closely at the countries associated with journals and countries where the work was done, we found the US published the most papers in both categories. However, many other countries do contribute greatly to experiments involving the constant – especially China, Russia, and UK in later years. Before 1938, the only countries contributing were the US, UK, and Germany. After 1938, many other countries began doing work on Newton's G while the US and UK continued to play prominent roles.

Many countries seem to regard work performed in their own country as the most important and have citation networks that show how often they cite their own work. Although international collaboration occurs, within this network, it is not seen in most of the papers. Popular articles from outside the country seem to be cited, but lesser known papers are only cited by the country that produced them. However, with the National Science Foundation's interest in the gravitational constant in 2016, more international collaboration is to be expected. This could help open the gateway for more experiments and for the sharing of information previously only well known within certain countries.

Further work with this citation network could include separating the network into smaller clusters based on the method used to measure G. Work with these clusters could help isolate different values of G and help determine if there is a relationship between values that use the same method of measurement.

Acknowledgements

We would like to thank Stephan Schlamminger for providing the papers on which this citation analysis is based. Without him, this paper would have been impossible to create. We would also like to thank the creators of Gephi, the citation analysis program, for creating an intuitive, well designed application for creating graphics.

# Appendix

**Figure 12.** A table containing the final list of sorted papers given by Schlamminger. Each number corresponds to a paper represented in Figures 1, 2, 7, 8, 9, 10, and 11.